\documentclass[usenatbib]{emulateapj}
\usepackage{amsmath}
\usepackage{graphicx}
\makeatother
\shorttitle{Magnetic fields in early protostellar disk formation}
\shortauthors{Gonz\'alez-Casanova et al.}

\begin{document}

\title{Magnetic fields in early protostellar disk formation}
\author{Diego F. Gonz\'alez-Casanova, and Alexander Lazarian}
%\altaffilmark{1}
\email{casanova@astro.wisc.edu}
\affil{Astronomy Department, University of Wisconsin-Madison, 475 North Charter Street, Madison, WI 53706-1582, USA}
\author{Reinaldo Santos-Lima}
\affil{Instituto de Astronomia, Geof\'isica e Ci\^encias Atmosf\'ericas, Universidade de S\~ao Paulo, R. do Mat\~ao, 1226, S\~ao Paulo, SP 05508-090, Brazil}
\begin{abstract} 
We consider formation of accretion disks from a realistically turbulent molecular gas using 3D MHD simulations. In particular, we analyze the effect of the fast turbulent reconnection described by the Lazarian \& Vishniac (1999) model for the removal of magnetic flux from a disk. With our numerical simulations we demonstrate how the fast reconnection enables protostellar disk formation resolving the so-called ``magnetic braking catastrophe''.  In particular, we provide a  detailed study of the dynamics of a 0.5~M$_\odot$ protostar and the formation of its disk for up to several thousands years.  We measure the evolution of the mass, angular momentum, magnetic field, and turbulence around the star. We consider effects of two processes that strongly affect the magnetic transfer of angular momentum, both of which are based on turbulent reconnection: the first, ``reconnection diffusion'', removes the magnetic flux from the disk; the other involves the change of the magnetic field's topology, but does not change the absolute value of the magnetic flux through the disk. We demonstrate that for the first mechanism, turbulence causes a magnetic flux transport outward from the inner disk to the ambient medium, thus decreasing the coupling of the disk to the ambient material. A similar effect is achieved  through the change of the magnetic field's topology from a split monopole configuration to a dipole configuration. We explore how both mechanisms prevent the catastrophic loss of disk angular momentum and compare both above turbulent reconnection mechanisms with alternative mechanisms from the literature.
\end{abstract}

\maketitle

\section{Introduction} 

Magnetic fields play an essential role for the formation of circumstellar disks. The modern theories of disk formation accept that magnetic fields dominate the dynamics in the disk \citep{li1996, li2011, li2013}.  The usual assumption behind this reasoning is based on the notion that in an ideal MHD situation, the magnetic field is frozen into the fluid. Hence, the magnetic field in the rotating disk is connected to the ambient interstellar gas, and this results in the loss of angular momentum of the infalling material. For the typical values of the interstellar magnetic fields, this magnetically-mediated loss of angular momentum may be so fast that accretion disks should not be able to form, a prediction grossly contradicted by ALMA observations \citep[e.g.][]{tobin2012}. This problem is so severe that it is known as the ``magnetic braking catastrophe''.

Naturally, it seems essential to account for the magnetic field diffusivity in a realistic interstellar environment. Non-ideal MHD effects can decouple matter from magnetic fields in different ways, allowing a diffusivity, different mechanisms for which have been discussed in the literature. The most common effect is related to ambipolar diffusion processes \citet{shu1983, shu2006}.  Ambipolar diffusion allows relative motions of the ionized gas and the neutral gas, and consequently allows magnetic fields to diffuse out of accretion disks.  However, to make the process work with sufficient efficiency, very special conditions are required, making ambipolar diffusion an unlikely candidate \citep[see][for a review]{li2014}. As a result, \citet{shu2006} suggests that an enhanced Ohmic dissipation may be present in the disks, but does not provide a quantitative physical mechanism to explain such an enhancement of Ohmic diffusivity. In this situation, the enhanced Ohmic dissipation cannot be considered as a general viable solution. Other plasma effects have also been considered: in particular, the influence of the Hall effect, were discussed, e.g. in \citet{krasnopolsky2010, braiding2012, tomida2013}. We discuss the relative importance of this effect further in the paper. There are other effects, such as the misalignment between the rotation axis of the disk and the magnetic field \citet{hennebelle2009, ciardi2010}. The aforementioned studies found that any misalignment has a considerable effect on the magnetic braking efficiency. This weakening, however, can only explain up to 50\% of the disk formation cases. 

In view of these difficulties, it is worth asking a question of whether magnetic fields are really frozen in the {\it realistic} interstellar medium, which is known to be turbulent. The evidence of turbulence in interstellar media is overwhelming \citep{elmegreen2004, mckee2007, falgarone2008, chepurnov2010}, and turbulence is known to change the transport properties of fluids, which begs the question of how turbulence can affect the evolution of magnetic fields in accretion disks. \citet[][henceforth LV99]{lazarian1999} presented a theory of fast 3D reconnection in turbulent media that suggested a possibility of magnetic flux having fast topology changes and diffusion in highly-conductive media. The LV99 model prediction of efficient motion relative to conducting fluid challenged the traditional concept of flux freezing of magnetic field in turbulent fluids. This notion was further elaborated and rigorously proven in later papers \citep[see][]{eyink2011,eyink2014}, while the theory and its major consequence to the violation of flux freezing were successfully tested in \citet{kowal2009, kowal2012, eyink2013}. These works constitute the justification for our further discussion of the dynamics of magnetic fields in turbulent fluids \citep[see also the recent review in][]{lazarian2015}.  

On the basis of the LV99 theory, \citet{lazarian2005} suggested that fast turbulent reconnection should substantially affect our understanding of the diffusion of magnetic fields at play during star formation \citep[see a recent review by][]{lazarian2014}. This may be important for the removal of magnetic field from molecular clouds and accretion disks and the redistribution of magnetic field in the interstellar medium. To stress the importance of reconnection and to distinguish the process from the commonly used ``ambipolar diffusion" process, the new process was termed ``reconnection diffusion''. In terms of magnetized disks, appealing to the data in \citet{shu2006}, \citet{lazarian2009} suggested that the process of reconnection diffusion could explain the problem of angular momentum of the accretion disks around young stars, thus resolving the magnetic braking catastrophe. A detailed discussion of the processes of reconnection diffusion and their relationship to star formation can be found in \citet{lazarian2012}. We stress that turbulence is assumed as a default interstellar condition: therefore, the diffusion mediated by turbulent reconnection may be called concisely ``reconnection diffusion'' rather than ``turbulent reconnection diffusion''. We also avoid using the term ``turbulent reconnection diffusion'' as it may be reminiscent of the poorly justified misleading concept ``turbulent resistivity'' that we also briefly discuss in the paper. 

The first numerical study of reconnection diffusion was performed in \citet{santos2010} with 3D MHD code and turbulent driving. This study convincingly demonstrated that reconnection diffusion can resolve a number of other paradoxes related to magnetic fields, e.g., the poor correlation of density and magnetic field in the diffuse interstellar medium. It also showed that magnetic fields can be quickly removed by reconnection diffusion during star formation activity. \citet{santos2012} subsequently focused on a rotationally-supported disk forming from turbulent medium, and provided numerical evidence that reconnection diffusion can efficiently solve the ``magnetic braking catastrophe''. 

Effects of turbulence on accretion disks have attracted the attention of other research groups. Soon after \citet{santos2012}, \citet{seifried2012, seifried2013} made similar simulations considering turbulent environments, but questioned the importance of reconnection diffusion effect in disks. The authors claimed---but did not observe---diffusion of magnetic field. Instead, they proposed that turbulence itself, without any effect from loss of magnetic flux, can solve the problem related to the disk's angular momentum. The subsequent work in \citet{santos2013} demonstrated that the diffusion of magnetic fields occurs at radii smaller than those considered by \citet{seifried2012}, and suggested that this may be the cause of the difference in conclusions between the two studies.

Due to the importance of the issue, we feel that additional, detailed studies of magnetic field and momentum diffusion are necessary.   In this paper, we quantify how the diffusion of the magnetic field depends on the turbulence properties of the medium, and we compare our results to those that follow from reconnection diffusion theory. We also test a claim in \citet{seifried2012, seifried2013} of whether the accretion disk formation is enabled through external spinning of the disk by turbulence. Furthermore, we present a second mechanism related to reconnection, a mechanism which can mitigate the problems of disk formation around stars.  In Section~\ref{sec:mdt}, we briefly explore magnetic reconnection theory, and how magnetic flux dynamics interact within turbulent fluids; in Section~\ref{sec:amt}, we explore the angular momentum transport theory; in Section~\ref{sec:num}, we describe the numerical code and setup for the simulations; in Section~\ref{sec:results}, we present our numerical results; in Section~\ref{sec:discussion}, we discuss our results and their correspondence with the theory; and in Section~\ref{sec:conclusion}, we give our conclusions.

%==========================================================
%==========================================================
\section{Magnetic Reconnection and Violation of Flux Freezing in Astrophysical Plasmas} \label{sec:mdt}

Astrophysical plasmas are highly conductive and therefore, it is usually assumed that any magnetic field diffusion is negligible on astrophysical scales. As a result, the concept of magnetic field freezing, based on the Alfv{\'e}n theorem, is widely used for describing the behavior of astrophysical magnetic fields \citep{alfven1942}. Nevertheless, nature demonstrates that, on many occasions (e.g., during solar flares), the conversion of magnetic energy into heat and energetic particles violates flux freezing \citep[see more more detailed discussion in][]{lazarian2015}.

Since most astrophysical fluids have high Reynolds numbers and demonstrate turbulence, it is natural to consider magnetic properties of turbulent fluids, as opposed to the laminar fluids considered within the context of the Alfv{\'e}n theorem. The theory of 3D turbulent reconnection (originally discussed in LV99, and summarized below) is the basis of our further consideration. 

\begin{figure}[t]
\centering\includegraphics[height=0.65\linewidth,clip=true]{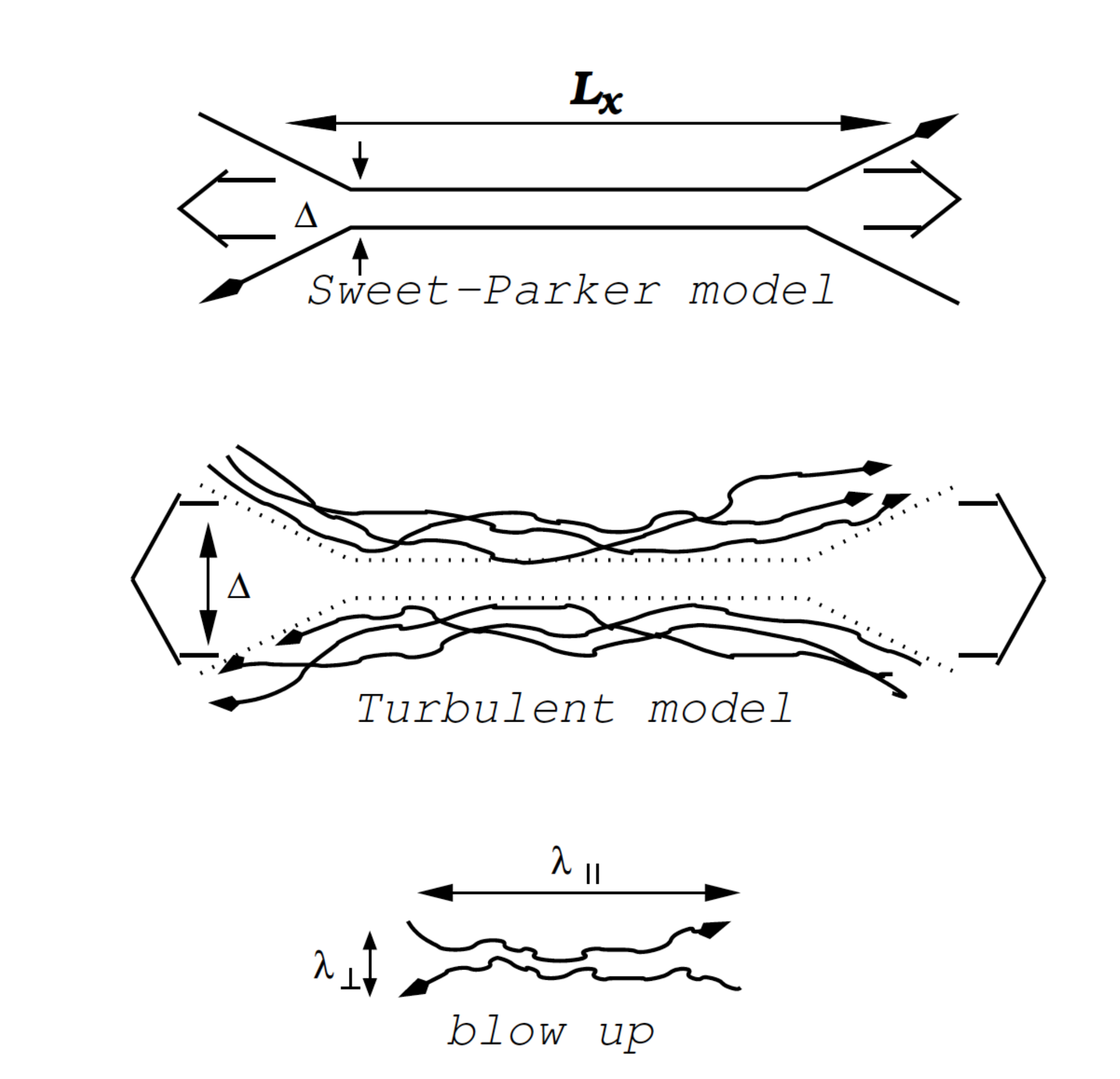}
\caption{{\it Top panel} The Sweet-Parker reconnection model. {\it Middle panel} The turbulent reconnection model (LV99). {\it Bottom panel} a blowup on scale of the turbulent reconnection model. $L_X$ is the scale of the reconnection and $\Delta$ the outflow thickness on the macroscopic level, $\lambda_\parallel$ and $\lambda_\bot$ at the individual scale on the turbulent model.}
\label{fig:theory}
\end{figure}

When one deals with simulations of astrophysical phenomena, it is essential to remember that---in most cases---a numerical experiment is a crude toy model of a complex astrophysical process. Additionally, numerical effects must be very carefully accounted for before any physical conclusion is made. This is absolutely essential for the processes of reconnection where the relevant dimensionless number that characterizes the Ohmic diffusivity (the Lundquist number) of fluids in real astrophysical system is orders of magnitude larger than anything that can be possibly achieved with simulations. The Lundquist number is: $S=L_x V_A/\nu$, where $L_x$ is the characteristic scale of the system, $V_A$ is Alfv\'{e}n velocity, and $\nu$ is Ohmic resistivity. Thus, the common argument of testing the accuracy of the results by changing the resolution by a small factor is very unreliable. Therefore, we claim that the effect of reconnection can be traced by simulations similar to \citet{santos2010, seifried2012, seifried2013} only if the physics of the reconnection diffusion is understood on a more fundamental level. In view of this, we stress that the reconnection diffusion that is based on the LV99 theory and its later extensions is not a concept obtained through brute-force experimenting with models of molecular clouds and accretion disks. The theory itself has been tested separately in specially-designed numerical experiments \citep{kowal2009, kowal2012, vishniac2012, eyink2013a, eyink2013}, and is being used as a foundation for the interpretation of our results.

We explain the previous point by stressing that magnetic field diffusion manifests in several ways in numerical simulations. Some diffusion is due to numerical effects, and is parasitic (i.e., it is not related to real astrophysical processes in high-conduction plasmas). However, this diffusion has been shown to be small, when compared with real diffusive-type processes induced by the violation of flux freezing in turbulent media \citep[see][and references therein]{lazarian2015}. The physical diffusion of the plasma and magnetic field is independent of resistivity. Therefore, even with limited numerical resolution, the large-scale slippage is physical in numerical simulations (i.e., it is not dominated by numerical effects). On this basis, we argue that the numerical results in this and earlier papers \citep[e.g.,][]{santos2013}, can be trusted. In other words, it is important to keep in mind that the present paper simply explores the consequences of the violation of flux freezing in turbulent media, while the concept is studied in detail elsewhere \citep[see][and references therein]{lazarian2015}.

The details of the reconnection theory in turbulent fluids have been discussed in a number a series of papers starting with LV99 (see \cite{lazarian2004, kowal2009, eyink2011, eyink2013}, see \cite{lazarian2015} for a review). Nevertheless, we believe that a simple illustration of the turbulent reconnection process may be appropriate. Figure \ref{fig:theory} illustrates the process of magnetic reconnection as suggested by LV99. Compared to the classical Sweet-Parker reconnection in laminar fluids, the outflow in the case of turbulence is limited not by the microscopic Ohmic diffusivity, but by turbulent magnetic field wandering. Mathematically, this means that the mass conservation constraint $v_{rec} L_x=v_A \Delta$ depends on the level of turbulence, e.g. for trans-Alfv{\'e}nic turbulence, $\Delta$ in turbulent fluids can be comparable with the scale $L_x$ at which magnetic fluxes come into contact. In this scenario, $L_x$ is the length of the magnetic sheet, $v_A$ the Alfv\'{e}n speed, $v_{rec}$ is the reconnection speed, and $\Delta$ is the size of the outflow region. 

The LV99 theory considers at all scales the turbulent movements of a magnetic field induced by the eddies of the host medium.  This changes the ejection scale of the reconnection in a way that does not depend on the Lundquist number. Hence, allowing a fast reconnection speed only depends on the properties of turbulence
\begin{equation}
v_{rec} = v_A(l/L_x)^{-1/2}(v_l/v_A)^2 \; ,
\end{equation}
where $l$ is the turbulence injection scale and $v_l$ is the velocity of the injection. This type of fast reconnection was shown in LV99 to make the \citet[][GS95]{goldreich1995} model of turbulence self-consistent. Without fast reconnection, the mixing motions associated with turbulent eddies rotating perpendicular to a magnetic field would result in the formation of unresolved magnetic knots. These unconstrained mixing motions induce the process of reconnection diffusion that we consider. 

We may mention parenthetically that if the reconnection were slow, any MHD turbulent numerical simulations would be meaningless. The huge differences in magnetic reconnection in astrophysical fluids for high Lundquist numbers and relatively low numerically-limited Lundquist numbers would make the turbulence in numerical simulations very dissimilar to its astrophysical counterpart.  

\begin{figure}[t]
\centering\includegraphics[height=0.65\linewidth,clip=true]{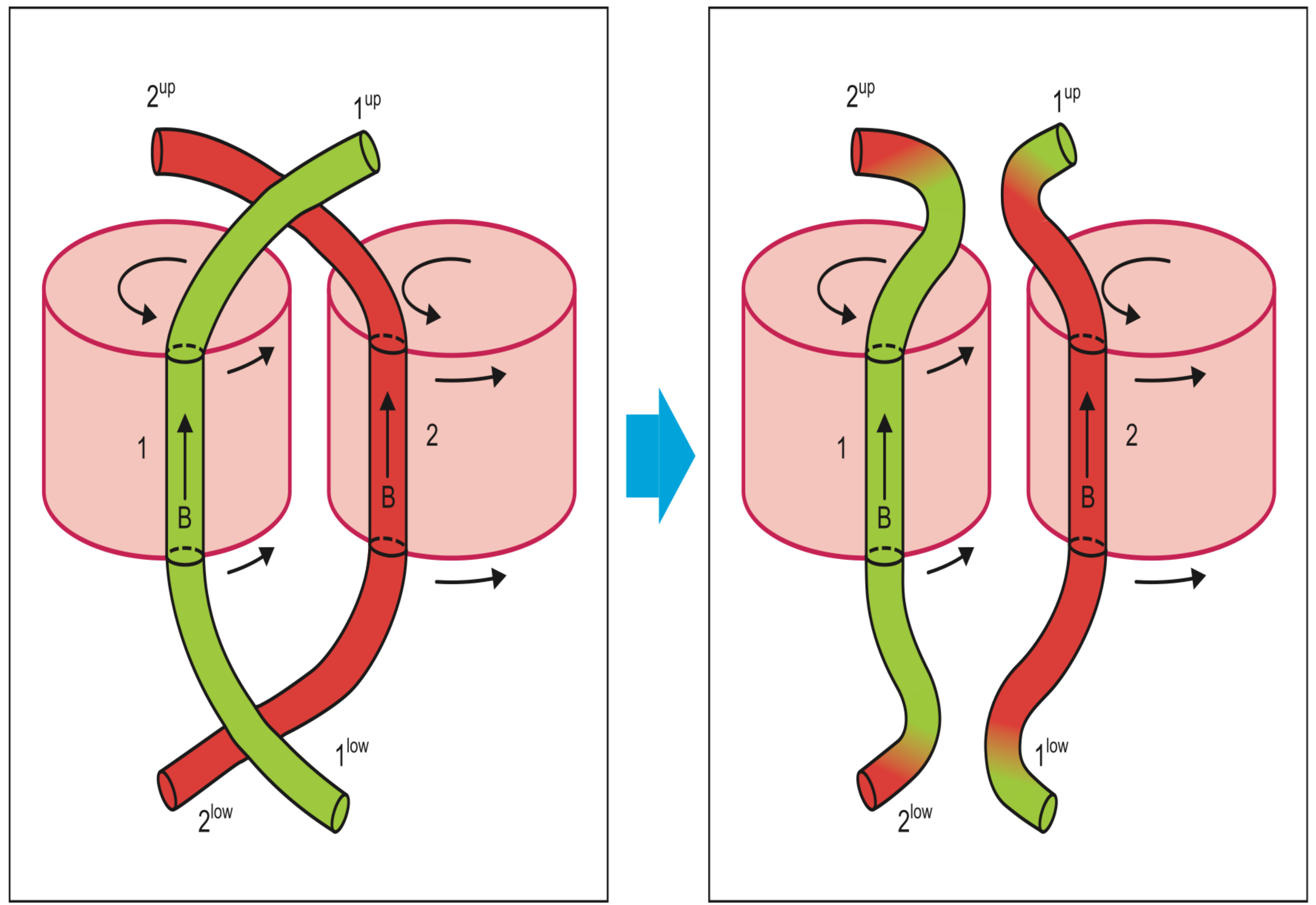}
\caption{The reconnection diffusion mechanism between to eddies rotates in the same direction.  Mixing of matter and the magnetic field occurs due to the reconnection of the magnetic field lines.}
\label{fig:lv}
\end{figure}

As discussed above, the ``magnetic braking catastrophe'' results from the assumption that magnetic field lines are frozen into plasmas. In the presence of magnetic reconnection, a magnetic field is constantly changing its topology and connections with time, and therefore, matter and the magnetic field move relative to one another. Figure \ref{fig:lv} illustrates reconnection diffusion at one scale as adjacent eddies exchange the parts of the flux tubes that are passing through them. One should keep in mind that such a process takes place at all scales of the turbulent cascade, enabling efficient exchange of plasma within magnetic flux tubes with different mass-to-flux ratios. A formal way to show the failure of flux freezing in turbulent media is demonstrated in \citet{lazarian2015}. A numerical study confirming the violation of flux freezing was performed with the extensive data set of MHD turbulence data in \citet{eyink2013a}.   

It is important to stress that reconnection diffusion is independent of the microscopic characteristics of the magnetized plasma, including the ionization state of the matter. Instead, it depends on the the scale and velocities of the turbulent eddies.  Flux transport by reconnection diffusion is faster where turbulence is stronger.  This point is crucial to understand the removal of magnetic fields in the theory of disk formation.  In the presence of gravity, such as in molecular clouds, diffusion will increase the segregation between the gas being pulled by the gravitational potential and the weightless magnetic field.

Finally, dealing with the theoretical foundations of the present work, we would like to remind our reader of a few facts about the turbulence statistics. Turbulence distributes kinetic energy over different scales, giving rise to an energy spectrum. In three dimensions, the energy spectrum for the Kolmogorov turbulence is: 
\begin{equation}
E(k) = C \varepsilon^{2/3} k^{-5/3} \; ,
\label{eq:powerlaw}
\end{equation}
where $C$ is a constant, $\varepsilon$ is the energy dissipation that depends on the viscosity, and $k$ is the modulus of the wave vector.  This power-law assumes a self-similarity at all the scales of the turbulence. The GS95 turbulence happens to be rather similar to the Kolmogorov turbulence in spite of the presence of magnetic field. The magnetic field makes the turbulence anisotropic, but the eddy motions perpendicular to the magnetic field preserve their eddy-type Kolmogorov nature \citep[see][for a review]{brandenburg2013}.

Because turbulence is a stochastic process, statistical tools are necessary to understand its behavior. In this paper, we will use structure functions, a statistical tool that uses a two-point correlation, analyzing the ``lag'' between the two points. In other words, the lag is the distance ($x_2-x_1$) between two points.  The lag $r$ is related to the wave number, ($k = |\mathbf{k}| \sim 2\pi/|\mathbf{r}|$).  The structure functions of order $n$ are:
\begin{equation}
SF(\mathbf{r}) = \langle [f(\mathbf{x_1}) - f(\mathbf{x_1}+\mathbf{r})]^n\rangle = \langle [f(\mathbf{x_1}) - f(\mathbf{x_2})]^n\rangle \; ,
\end{equation}
were $f$ can be any function. In the case of turbulent motions, $f$ is the turbulent velocity. However, the energy spectrum of turbulence is obtained from second-order structure functions (equation \ref{eq:powerlaw}).  

Due to its construction, the amplitude of a structure function is related to the intensity of the turbulence. Therefore, a correlation between the magnetic flux diffusion and the amplitude of the structure function is expected.  This is important because, as pointed out in \citet{santos2013}, the mass-to-flux ratio $\mu$ cannot be used as a quantity to analyze the correlation between diffusion processes and turbulence.

\section{Angular momentum transport during disk formation} \label{sec:amt}

\subsection{General considerations}

In the paper, we study the angular momentum flux in the accretion disk.  The angular momentum flux has both components associated with plasmas and and a magnetic field.  In general, the angular momentum follows the relation for a constant $r$:
\begin{equation}
\label{eq:amf}
\partial_t (\rho \epsilon_{ijk} r_j u_k) + \partial_l (\Lambda_{il}) = \epsilon_{ijk} r_j f_k \;,
\end{equation}
where $\rho$ is the density, $f$ external forces such as turbulence injection, $\epsilon_{ijk}$ the Levi-Civita symbol, and  $\Lambda_{kl}$ is the angular momentum tensor.  The angular momentum tensor is construct using the momentum tensor including viscosity and the magnetic component. For cartesian coordinates the momentum tensor is: $\Pi_{kl}$ = $p \delta_{kl}+\rho u_k u_l + \nu (\partial_i u_k + \partial_k u_i-2/3\partial_j v_j \delta_{ik})$ and the angular momentum tensor is: $\Lambda_{kl} = \epsilon_{kij} \rho r_i (u_j u_l - \nu (\partial_j u_l + \partial_j u_l)) + \epsilon_{kij} r_i B_j B_l = \Lambda_b + \Lambda_m$, where $r$ is the position of the element, $u$ the speed, $B$ the magnetic field, $\Lambda_g$ is the matter component of the momentum tensor, and $\Lambda_b$ the magnetic one. The angular momentum flux to the disk is then given by the radial and $z$ component of $\Lambda$.  This tool permits a measurement of the direction and magnitude of the angular momentum flux going in or out of the disk that is evaluated through time.  

Consider a disk being built up around a protostar. Using a cylindrical  coordinate system $(R, \theta, z)$ with its origin in the protostar and the z-axis alined in the direction of the rotation  axis of the disk, the equation describing the evolution of the the $z$ component  of the angular momentum density $m_z \equiv \rho R u_{\theta}$ is given by:
\begin{equation}
	\frac{\partial m_z}{\partial t} = 
	- \mathbf{\nabla} \cdot \left( m_z \mathbf{u} - R B_{\theta} \mathbf{B} \right) 
	- \frac{\partial p}{\partial \theta} + \frac{\partial \Phi}{\partial \theta}+ R (\mathbf{\nabla} \cdot \mathbf{\mathbf{\sigma}})_\theta \;,
%-\rho R \frac{\partial \Phi}{\partial \theta} 
\end{equation}
where the viscosity is included in the stress tensor $\sigma$ and $\Phi$ is the gravitational potential. Let us assume the idealized case of axisymmetry in what follows, dropping the terms related to the thermal pressure and gravitational acceleration. Consider a control volume of cylindrical shape, with radius $R_0$ and half-height $z_0$ (centered on the origin and aligned with the coordinate system). The rate of change of the total `z' component of the angular momentum ($J_z$, henceforth just angular momentum) of the gas inside the volume is given by:
\begin{equation}
\frac{d J_z}{dt} = \tau_{gas} + \tau_{B \;},
\label{eq:torque1}	
\end{equation}
where $J_z = 2 \pi R_0^2 z_0 \langle m_z \rangle_V$ (the brackets $\langle \cdot \rangle_V$ mean 
an average over the cylindrical volume), and the gas and magnetic torques are given respectively by:
\begin{eqnarray}
	\tau_{gas}&=&\pi R_0^2 \left( \langle m_z u_z \rangle_{S_2} - \langle m_z u_z \rangle_{S_1} \right) \nonumber \\
			 &&- 4 \pi R_0 z_0 \langle m_z u_R \rangle_{S_3} \;, 
\end{eqnarray}
\begin{eqnarray}
	\tau_{B}&=&\pi R_0^2 \left( \langle R B_{\theta} B_z \rangle_{S_1} - \langle R B_{\theta} B_z \rangle_{S_2} \right) \nonumber \\
	&&+ 4 \pi R_0 z_0 \langle R B_{\theta} B_R \rangle_{S_3} \;, 
\label{eq:torque}
\end{eqnarray}
where $\langle \cdot \rangle_{S_1,S_2,S_3}$ mean, respectively, averages over the upper, bottom, and lateral surfaces of the cylinder.

The matter torque $\tau_{matter}$ is due to the flow of matter carrying angular momentum into and out of the disk as a whole. Of course it must be predominantly positive during the disk formation. The magnetic field torque $\tau_B$, on the other hand, must be negative, as it is explained below. 

The $B_{\theta}$ component has its origin from two sources. One is the vertical shear of the rotation velocity at the disk  upper and bottom surfaces, which stretches the vertical component $B_z$ at expenses of the rotation. This disturbance of the field lines propagates out of the disk, generating a helical pattern of the field lines, which can be described as torsional Alfv\'en waves carrying angular momentum \citep[see][]{joos2013}. The other source is the shear of the rotation velocity in the radial direction, caused by the differential rotation inside the disk, which stretches the radial component $B_R$ and transfers angular momentum to external radius. 

Therefore, the negative torque produced by the magnetic field lines transports angular momentum from the disk to the envelope. \citet{mouschovias1977} calculated the time-scale for the braking of the disk rotation, when the  mean magnetic field is aligned to the rotation axis of the disk:
\begin{equation}
\tau_{\parallel} = \frac{\rho_{d}}{\rho_{e}} \frac{Z_d}{v_{A,e}} \equiv \left( \frac{\pi}{\rho_{e}} \right)^{1/2} \frac{M_d}{\Phi_B} \; ,
\end{equation}
where $\rho_{d}$ and $\rho_{e}$ are the densities of the disk and envelope, respectively, $Z_d$ is the half-thickness of the disk, $v_{A,e}$ is the Alfv\'en speed in the envelope, $M_d$ is the disk mass, and $\Phi_B$ is its magnetic flux.  Using realistic values of magnetic field in cloud cores, it is found that in the absence of turbulence the associated timescale is short, when compared to the timescale for the formation of the disk. This process has been demonstrated numerically in several studies e.g. \citep{krasnopolsky2010, santos2012}.

\subsection{Numerical studies of the magnetic flux loss}

Using 2D, high resolution numerical simulations, \citet{krasnopolsky2010} demonstrated that a rotationally supported disk can be formed when resistivity enhanced orders of magnitude compared with the expected resistivity is present. This artificially high resistivity is decoupling magnetic fields from the disk material. He found the magnetic diffusivity is required to be $\sim 10^{20}$~cm$^2$~s$^{-1}$ for the realistic values of magnetic field in the initial cloud. For realistic Ohmic resistivity values, the diffusivity values are significantly below this estimate (e.g. $\sim 10^{17}$~cm$^2$~s$^{-1}$ for a density of $10^{-13}$~g~cm$^{-3}$), signifying an effective magnetic braking, and preventing the formation of a Keplerian disk. Not surprisingly, they found that a rotationally-supported disk formed in the presence of the enhanced resistivity has a substantially reduced magnetic flux. Therefore, the magnetic torque is mitigated due to the reduction of $B_z$. 

As we discussed earlier, an efficient magnetic diffusivity happens naturally in turbulent plasmas and does not require any enhancement of Ohmic effects. In fact, \citet{santos2012} analyzed numerically the effect of reconnection diffusion \citep[]{lazarian2005} as a physical mechanism that is able to provide the magnetic diffusivity at the levels required in \citet{krasnopolsky2010}. Using 3D simulations and initial conditions analogous of those in \citet{krasnopolsky2010}, they demonstrated that  when the turbulent molecular cloud undergoes star formation, a rotationally-supported disk forms without the need for enhanced Ohmic resistivity. They employed turbulence injection with features needed to provide the levels of magnetic diffusivity required ($\eta_{turb} \sim 10^{21}$ cm$^2$s$^{-1}$, ~\footnote{This estimate  for the reconnection diffusion  is valid for super-Alfv\'{e}nic turbulence (see more details in \citet{eyink2011}.)}). They demonstrated the newly-formed disk has a smaller magnetic flux than the pseudo-disk formed when turbulence is absent  (that is, when only numerical resistivity is present and the magnetic braking is efficient), but this magnetic flux is larger than that from the  disk formed without turbulence but in the presence of super-resistivity \citep[see details in][]{santos2012, santos2013}.  Appealing to the theory of turbulent reconnection, the authors explained that the observed effect is real (i.e., it is not related to limited numerical resolution). This studies are the starting point of our present research. 

As we mentioned earlier, the conclusion that reconnection diffusion (RD) dominates flux loss in turbulent disks was challenged in \citet{seifried2013} who also studied the formation of protostellar disks inside a turbulent molecular cloud. They followed the stellar formation process since the molecular cloud scales down to the scales of the protostellar disks, employing in their 3D numerical simulations the techniques of Adaptive Mesh Refinement and sink particles \citep[see][]{federrath2010}, solving the self-gravity in a self-consistent way. A turbulence spectrum is present in the initial cloud, and naturally develops as time passes due to cascading, without need for driving. They also demonstrated the formation of protostellar disks sustained by rotation when the turbulence was present, while the Keplerian disks failed to form in the absence of turbulence. 

For a set of rotationally-sustained disks formed from their simulation, \citet{seifried2013} measured the ratio of two timescales, the disk formation torque and the magnetic braking torque, i.e. $|\tau_{matter} / \tau_{B}|$ (inside cylindrical volumes of radius varying from $R_0 = 1-1000$~AU and half-height $z_0 = 40$~AU). They found this ratio to be $< 1$ in their simulation without turbulence (for the interval $7 < R_0 < 50$), as one would expect when the magnetic braking is efficient in extracting angular momentum from the disk. For their simulations where turbulence was present, they found the ratio $|\tau_{matter} / \tau_{B}|$  to be larger than $1$ ($\sim 100$ at $R_0 \sim 5$ and decaying to $\sim 3$ at $R_0 \sim 100$~AU for most of models; for one of their models, this ratio starts from $\sim 1000$ at $R_0 \sim 5$ and decays to $\sim 100$ for radius larger than $100$~AU; see Figure 4 in \cite{seifried2013}). The authors did not observe the decrease of the magnetic flux due to reconnection diffusion. Instead one of their interpretations of the results was that the turbulence {\it decreases} $\tau_{matter}$ due to the turbulent shear local, which feeds the angular momentum of the disk. The present study is aimed to provide the diagnostics to analyze this effect versus the effect of magnetic reconnection.

The mass-to-flux ratio is a tool used to measure the formation of the disk.  As \citet{santos2012} showed, this tool can be misleading.  Nonetheless, the trend is an increase of the mass-to-flux ratio over the disk formation, implying a reduction of $B_z$.  In the case where the misalignment between the magnetic field and angular momentum is small, the mass-to-flux ratio trend can only be explained by RD in the presence of turbulence.  This effect has been measured in different studies \citep{santos2012, joos2013, seifried2013}.  In particular, \citet{joos2013} measured it in the disk scale and found that the increase in the mass-to-flux ratio can be avoided if the turbulence is injected after the initial phase of the disk formation. Therefore, in order to observe the effects of turbulence in the  envelope---excluding the increase in the mass-to-flux ratio---turbulence cannot be present at the beginning of the simulation.

%=======================================
\section{Numerical Setup}\label{sec:num}

We use the {\it AMUN} code, a 3D Cartesian Godunov MHD code \citep{kowal2007, falceta2008}.  The code solves the resistive MHD equation considering an isothermal equation of state:

\begin{eqnarray}
\frac{\partial \rho}{\partial t} + \mathbf{\nabla} \cdot (\rho \mathbf{v}) = 0 \;, \hfill \nonumber \\
\rho \big( \frac{\partial }{\partial t} + \mathbf{v} \cdot \mathbf{\nabla} \big) \mathbf{v} = -c^2_s \mathbf{\nabla} \rho + (\mathbf{\nabla} \times \mathbf{B}) \times \mathbf{B} - \rho \mathbf{\nabla} \phi + \mathbf{f} \;, \hfill \nonumber \\
\frac{\partial \mathbf{B}}{\partial t} = \mathbf{\nabla} \times (\mathbf{v} \times \mathbf{B}) +\eta \mathbf{\nabla}^2 \mathbf{B} \;, \hfill \nonumber \\
 \mathbf{\nabla} \cdot \mathbf{B} = 0 \hfill \;, \nonumber\\
\end{eqnarray}

where $\rho$ is the density, $\mathbf{v}$ is the velocity, $c_s$ is the sound speed, $\mathbf{B}$ is the magnetic field, $\phi$ is the gravitational potential, $\eta$ is the resistivity, and $\mathbf{f}$ is an external random force. The code uses a one-fluid approximation, which is valid since reconnection diffusion is not highly dependent on the ionization state. For this case, $f$ is a random force that injects the turbulence. The turbulence is distributed isotropically in Fourier space, as governed by a Gaussian distribution, centered around the injection scale.  The injection scale is defined as: $2.5~l_{inj} = l_{box}$. The turbulence is driven during the first half of the simulation, since star formation regions do not have constant turbulence injection. The turbulence velocity is $\sim 8 \times 10^{4}$~cm~s$^{-1}$.

The AMUN code uses second-order, shock-capturing Godunov and time-evolution Runge-Kutta approximations.  The fluxes are calculated by an HLL solver.  Open boundary conditions with null magnetic potential were used.  This condition prevents artificial artifacts such as {\it spiral arms} and {\it corners} in the disk \citep{santos2012}. The gravitational potential only considers one sink particle, the protostar. Since there is no self-gravity, no sink particles are created.  The accretion technique is described in \citep{federrath2010}, and assumes a smoothing radius (inside which angular momentum and mass are conserved) to solve the potential well.

\subsection{Protostellar Disk model}

The initial setup is a collapsing cloud progenitor with initial constant rotation. The velocity profile is given by  $v_\phi = c_s \rm{tanh}(R/R_s)$, where the maximum velocity is $\sim 2 \times 10^{4}$~cm~s$^{-1}$ and $R_s \sim$200~AU, as studied by \citet{krasnopolsky2010}. The magnetic field is uniformly distributed with a magnitude of $35~\mu G$ in the $z$ direction. The protostar has an initial mass of 0.5~M$_\odot$. 
\begin{figure}[h]
\centering\includegraphics[width=\linewidth,clip=true]{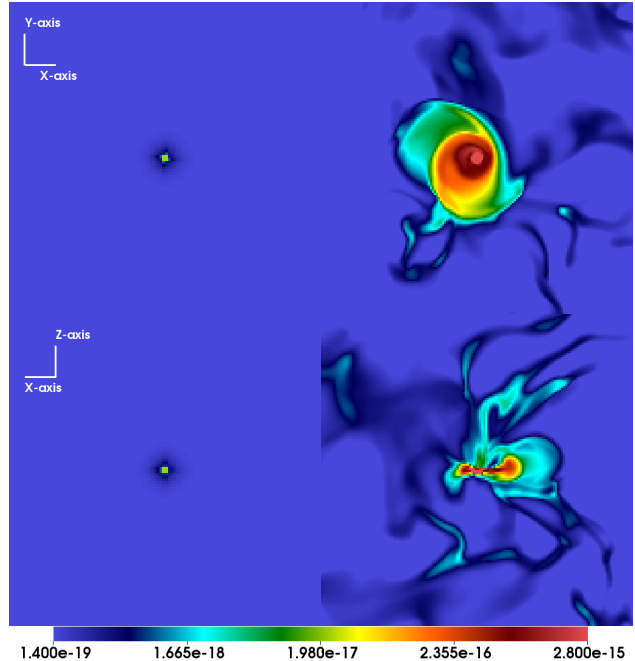}
\caption{Density profile in log scale of the disk $[g~cm^{-3}]$.  {\it Left panels}: $t_1=0$~yr and the {\it right} size at $t_2=28,538$~yr. {\it Top panels}: ``XY'' plane at $z=0$. {\it Bottom panels}: a ``XZ'' plane at $y=0$. } 
\label{fig2}
\end{figure}
 
The simulations were made in a $512^3$ grid. The initial setup is shown in Figure \ref{fig2}.  Based on LV99 reconnection theory, we know that reconnection diffusion does not depend on the resistivity---and therefore, on the resolution. Simulations with lower resolution were made confirming, the no dependance with resolution.  This has also been observed in numerical works by \citet{santos2010}, although as we discussed earlier, such resolution studies would not be convincing in the absence of the reconnection theory, which is the basis of the RD. More over, in order to test that is RD the diffusion mechanism, non-turbulent models were implemented showing that just numerical diffusion is not sufficient to form a stable disk. The minimum scale of the grid is $\sim$7.3~AU, and so the total physical size for the simulation is $\sim$4010~AU.  The simulation evolves for 28,538~yr.

%============================================
\section{Results} \label{sec:results}

\begin{figure}[h]
\centering\includegraphics{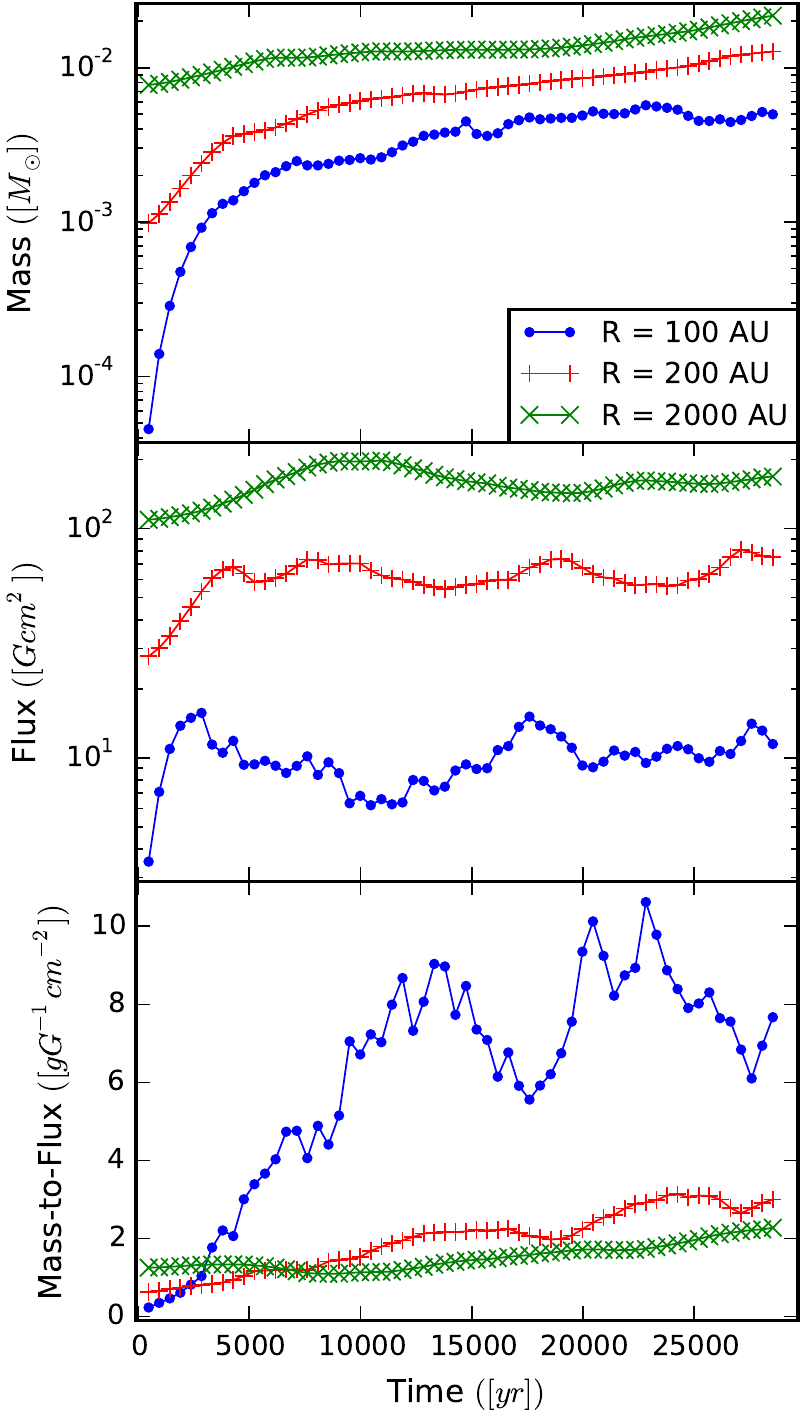}%[width=\linewidth,clip=true]
\caption{ Time evolution of the mass $M$ ({\it top panel}), magnetic-flux $\Phi$ ({\it middle panel}), 
and mass-to-flux ratio $M/\Phi$ ({\it bottom panel}) inside spheres of different radius $r$ 
centered at the protostar. {\it Blue lines}: $r = 100$~AU; {\it red lines}: $r = 200$~AU; 
{\it green lines}: $r = 2,000$~AU. The magnetic flux was calculated using the 
average $z$-component of the magnetic field inside the sphere ($\Phi = \pi r^2 \langle B_z \rangle$.)} 
\label{fig:santos}
\end{figure}

We first corroborate and extend the results of \citet{santos2012}, analyzing the magnetic flux $\Phi$, the mass-to-flux ratio $M/\Phi$, and the mass $M$ through spheres of different radius $R$, centered at the protostar. The magnetic flux $\Phi$ is calculated using the average of the $z$-component of the magnetic field inside the sphere ($\langle B_z \rangle$): $\Phi = \pi R^2 \langle B_z \rangle$. The results are shown in Figure \ref{fig:santos}.

Results shown in Figure \ref{fig:santos} confirm the findings obtained in the case of turbulent formation of disk in \citet{santos2013} but with a higher numerical resolution than in the aforementioned work. Lower panel of Figure \ref{fig:santos} shows the mass-to-flux ratio of the disk. The general trend of the mass-to-flux ratio is to increase with time,  albeit with small bumps, caused by either the change in the magnetic field topology  (see Section 5.2) or the turbulent movements of the medium. 

To understand the role of the reconnection diffusion during the disk formation, 
we analyze quantitatively the evolution of the:  
\begin{enumerate}
\item Magnetic field and angular momentum of the disk;
\item Turbulence statistics inside the disk;
\item Topology inside the disk-envelop system.
\end{enumerate}

The following analysis consider the evolution of several physical quantities integrated inside the volume of cylinders with different radii $R$ around the protostar, unless otherwise is stated. We employ cylinders with radii of 47, 94, 141 and 188~AU  (equivalent to 6, 12, 18 and 24 grid cells, respectively). All the cylinders have height of 188~AU (or 24 grid cells).

\subsection{Disk time evolution properties}

\begin{figure}[]
\centering\includegraphics[width=\linewidth,clip=true]{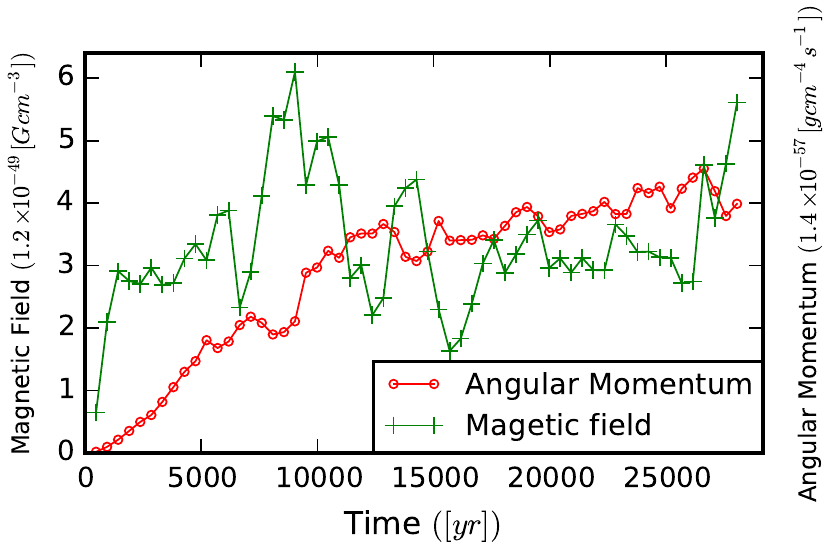}
\caption{Time-evolution of the volume-averaged magnetic field $\langle B \rangle$ ({\it green}) and the angular 
momentum density $\langle m \rangle = \langle \rho R u \rangle$ ({\it red}) inside the cylinder with radius $R=47$~AU. 
The same shape holds for the cylinders with larger radius.}
\label{fig:evol}
\end{figure}

\begin{figure}[t]
\centering\includegraphics[height=0.65\linewidth,clip=true]{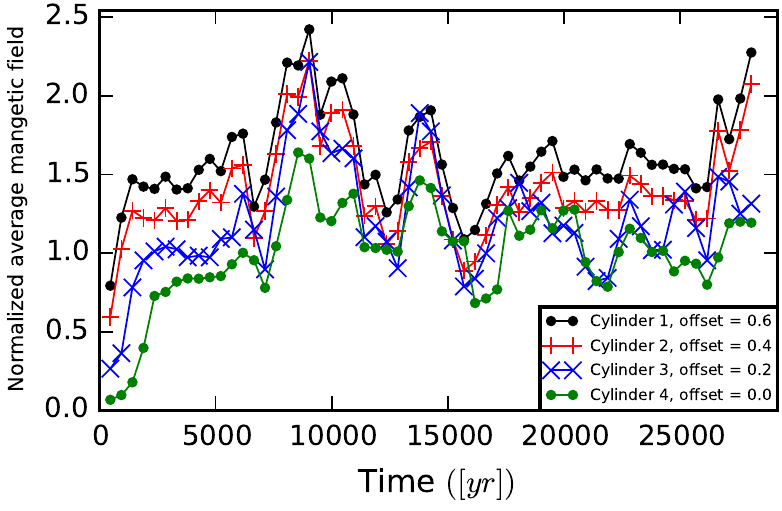}
\caption{Evolution of the volume-averaged of the angular momentum in cylinders of different radius $R$.  All the values are normalized by it respective time average at each cylinder and then off set by `0.4' from the largest cylinder. {\it Black:} $R=47$~AU;{\it red:} $R=94$~AU; {\it blue:} $R=141$~AU; {\it green:} $R=188$~AU.}
\label{fig:magevol}
\end{figure}

We follow the time-evolution of the angular momentum density $m = \rho R u$ and of the mean magnetic field $|B|$ averaged by the volume inside each cylinder (see Figures~\ref{fig:evol} and~\ref{fig:magevol}). Both magnetic field and angular momentum grow during the first stages, when the disk is forming. At $\approx 7,600$~yr, reaches a peak, decreasing afterwards finding a fluctuating steady point. Meanwhile, the angular momentum increases monotonically (within the precision allowed by turbulent effects), up to a steady state, allowing a rotationally-supported disk 
\citep{santos2012}.  

Aiming to compare the torque exerted by the magnetic field lines with that exerted by the gas on the angular momentum of the disk, we integrate the radial and vertical components of the angular momentum flux through the surfaces of the cylinders with different radii (Equation \ref{eq:amf}) The gas and magnetic field components of the angular momentum flux tensor (respectively $(\Lambda_{ij})_m$ and $(\Lambda_{ij})_b$) 
are analyzed separately. The angular momentum fluxes in the radial direction are given by $\Lambda_m = -\rho z (u_R u_{\theta} + \nu (\partial_{\theta} u_R + \partial_R u_{\theta}))$, and  $\Lambda_b =  z B_R B_{\theta}$. In the radial component $\Lambda_m = -\rho R (u_z u_{\theta} + \nu (\partial_{\theta} u_z + \partial_z u_{\theta}))$, and $\Lambda_b =  R B_z B_{\theta}$.

% \begin{equation}
% \left( \Lambda_{zR} \right)_m = \rho R (u_R u_{\theta} + \nu (\partial_{\theta} u_R + \partial_R u_{\theta})), 
% \end{equation}
% \begin{equation}
% \left( \Lambda_{zR} \right)_b = - R B_R B_{\theta},
% \end{equation}
% while the fluxes of $m_z$ in the $z$-direction are given by:
% \begin{equation}
% \left( \Lambda_{zz} \right)_m = \rho R u_{\theta} u_z + (COMPLETE \;\;\; VISCOUS \;\;\; COMPONENT), 
% \end{equation}
% \begin{equation}
% \left( \Lambda_{zz} \right)_b = - R B_{\theta} B_z.
% \end{equation}

\begin{figure}[h]
\centering\includegraphics{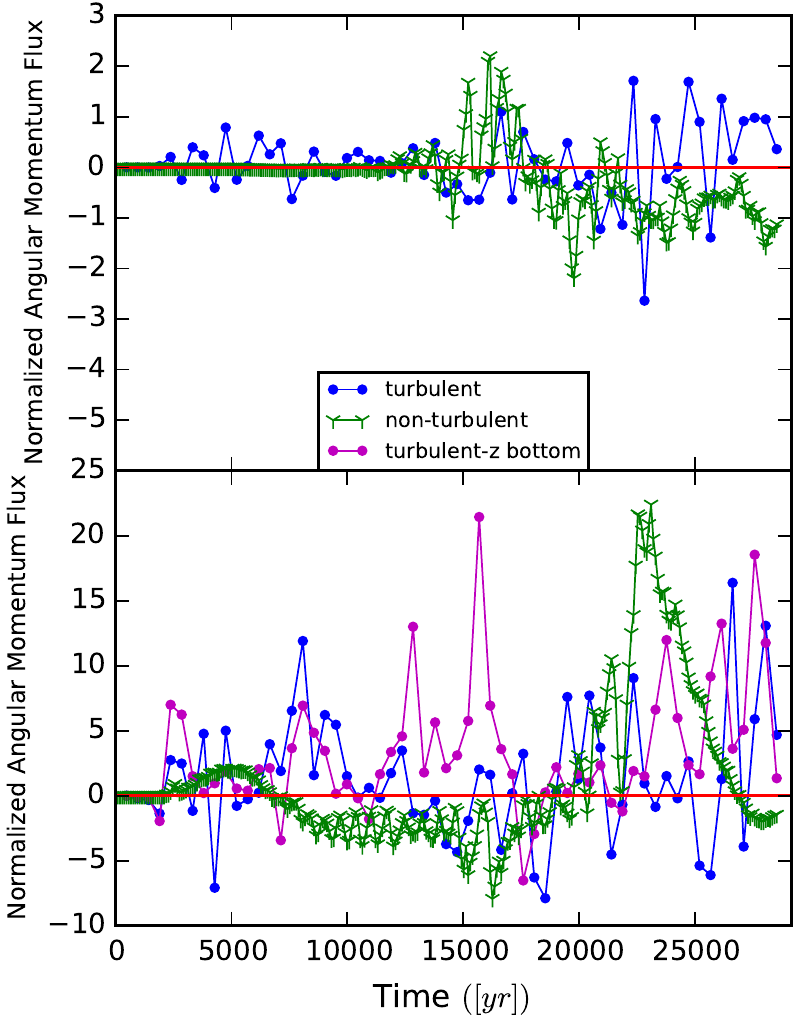}%[width=\linewidth,clip=true]
\caption{Flux of the angular momentum (normalized) integrated over the cylinder with radius  R=47~AU, for the simulations with ({\it blue lines}) and without ({\it green lines}) turbulence injection. {\it Top panel:} flux contribution through the $R$ (radial) direction.{\it Bottom panel:} flux contribution through the upper cylindrical surface ($z$-direction); in magenta it is shown the flux contribution through the bottom cylindrical surface.The quantities are normalized by a factor of $10^{52}~g~s^{-2}~cm$for the turbulent case (blue lines) and $10^{53}~g~s^{-2}~cm$ for the non-turbulent one (green lines).}
\label{fig:flux}
\end{figure}

The effective viscosity $\nu$ due to numerical effects in the present simulations has a value estimated of $\sim 3 \time 10^{18}$~cm$^2$~s$^{-1}$. The integral of $\Lambda$ over the desired surface gives the angular momentum flux into or out of the surface (Figure~\ref{fig:flux}). The non-turbulent model has the same initial setup as the turbulent one, but with no turbulence injected into the simulation. A positive value on the flux implies an outflow, while the inflow has a negative value. The sum of the fluxes over all the surfaces delimiting the cylindrical volume gives the negative of the torque exerted in the $z$ direction (equation \ref{eq:torque1}).

There are three distinct phases for the angular momentum flow in the radial and $z$ direction. The $z$ direction has a similar contribution from both upper and bottom cylindrical surfaces.  The first phase is characterized by an inflow of angular momentum in the radial direction and outflow of angular momentum in the $z$-direction (corresponding to the buildup of the magnetic flux and angular momentum of the disk).  The second phase has an outflow/steady state of the magnetic field contribution to the angular momentum flux. The third phase has an in- or outflow of angular momentum flux that corresponds to the steady-state in the disk's magnetic field.The magnetic field contribution due the magnetic field torque ($\tau_{B}$) is of the order of $10^{52}~g~s^{-2}~cm$, while the contribution coming from the gas torque $\tau_{gas}$ is one order of magnitude smaller. For both models (with and without turbulence) the torque caused by the gas flow (including viscosity) is smaller than the torque exerted by the magnetic field. 

The first thing we find is that the angular momentum flux due to the magnetic field, $\Lambda_b$, is the one that dominates the flux to the disk (Figure~\ref{fig:flux}). Therefore the feeding of angular momentum into the disk caused by the inflow of gas (referred as spin-up of the disk due to the local shear in \citet{seifried2013}) could not equilibrate the extraction of angular momentum by the magnetic torque in our simulation.
%(DIEGO, A CLOSE COMPARISON WITH THE RESULT ON THE RATIO BETWEEN THE TORQUES OBTAINED IN SIEFRIED ET AL 2013 IS CONVENIENT HERE)} 

The torque exerted on the disk for the model without turbulence has fluctuations on the second half of the simulation (Figure ~\ref{fig:flux}). 
%{\color{red} (EXPLAIN BRIEFLY WHY)}
These oscillations are a product of the magnetic braking. The breaking of the disk produces variations to the mass, velocity and inclination angle that account for the oscillation on the angular momentum fluxes in all the surfaces.  

\begin{figure}[h]
\centering\includegraphics[width=\linewidth,clip=true]{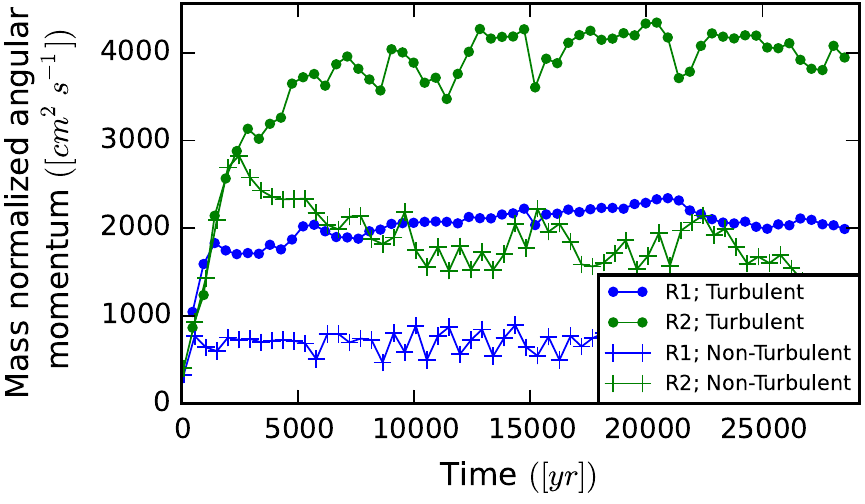}
\caption{ Evolution of the mass-normalized angular momentum ($J_z/M$) 
inside the cylinders of radius R1=47~AU ({\it blue lines}) and R2=94~AU({\it gree lines}) for the models 
with and without injection of turbulence.}
\label{fig:flux-ntt}
\end{figure}

The second analysis done to understand the role of the turbulence in the angular momentum transfer is to calculate the total angular momentum of the disk normalized by its mass. We compare this analysis for both 
the turbulent and non-turbulent models (Figure~\ref{fig:flux-ntt}). While the mass-normalized angular momentum ($J_z/M$;$J_z=\int_V m_z dV $) in both cases is similar ---implying that the shearing of the turbulent movements do not increase the angular momentum --- the total angular momentum of the disk is different for the two cases.

At the smallest disk the momentum is similar, i.e. the velocity profile for the disk remains Keplerian, the disruption of the disk does not affect this scale (Figure \ref{fig:flux-ntt}).  At the second disk for the turbulent model the angular momentum arrives to a stable Keplerian configuration due to the disk formation.  In the non-turbulent model the momentum decreases due to the disk disruption and the lost of the disk velocity. 

\subsection{Statistical properties of the turbulence}

\begin{figure}
\centering\includegraphics[width=\linewidth,clip=true]{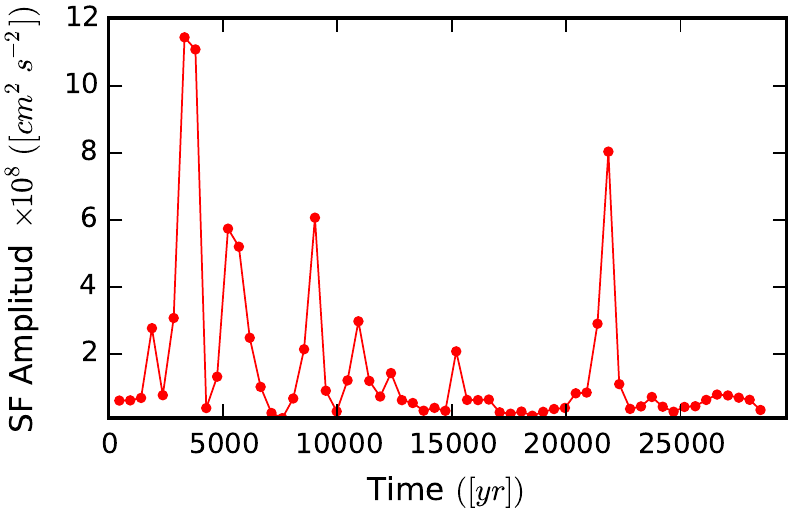}
\caption{ Intensity of the structure function for the coherent velocity (the velocity field with out the rotation) inside the cylinder with radius R=$94$~AU. The lag used is 47~AU (6 cells), averaging over 10 points}. 
\label{fig:sf}
\end{figure}

Turbulence plays an important role in disk formation. Therefore, it is important to 
understand its features in our simulation.  In order to characterize the turbulence intensity in the disk region,
we use second-order structure functions of the velocity on an annulus in the $z=0$ plane, at the same radii used 
by the cylinders employed in the volume analysis in the last subsection. The velocities use for the structure functions do not take into account the rotation and the inflow velocity. The time-evolution of the structure function of the velocity field for a fixed lag, representing the squared of the turbulent velocity in the scale of the lag is shown in Figure~\ref{fig:sf}. This tool allows us to understand the evolution properties of the turbulence, in relation to the angular momentum flux.

Fixing an annulus and a time, the turbulence spectrum can be calculated by varying the lag of the structure functions.
We found that the energy power spectrum follows a power law with a slope of $\approx -1.5$ for the scales that range form 47 AU to 2000 AU. This power spectrum is closer to the Kraichnan index \citep{brandenburg2013}. The difference in the power index could be due to the low spatial resolution due to more extended bottleneck of 
the MHD simulations compared to its hydrodynamic counterpart \citep{beresnyak2010}. 

The intensity of the turbulence (Figure~\ref{fig:sf}) in a magnetized medium is related to the magnetic reconnection and therefore to the process of reconnection diffusion \citep[see][for a review]{lazarian2014}. In particular the diffusion is related to the angular momentum flux going into the disk  (Figure~\ref{fig:flux}).  There is some relation form the properties of the angular momentum flux in the radial direction and the intensity of the turbulence.

\subsection{Topology of the magnetic field}

\begin{figure}[]
\centering\includegraphics[width=\linewidth,clip=true]{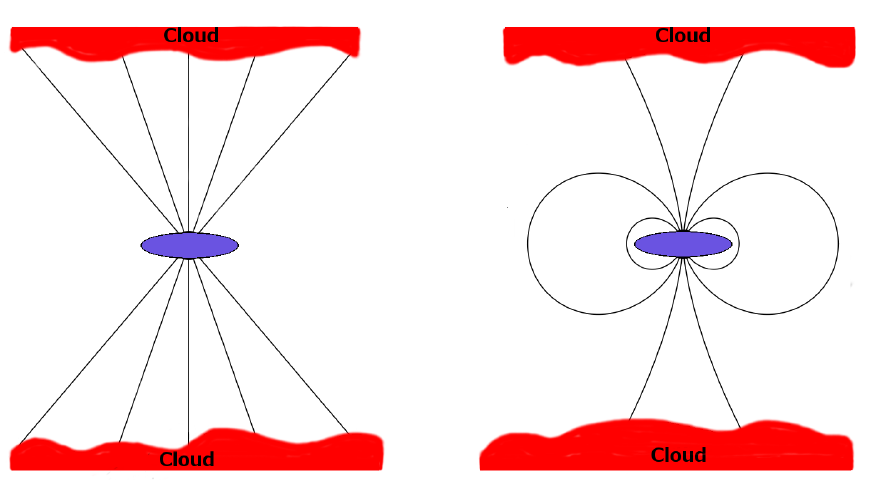}
\caption{Diagram representing the split monopole and dipole magnetic field configuration and their interactions with the cloud.  The magnetic field in the split-monopole configuration has more interactions with the ambient (cloud) magnetic field than the dipole configuration.}
\label{fig:diag}
\end{figure}

In addition to the transport of magnetic flux during the formation of the protostellar disk we expect another effect related to turbulent reconnection to become important. A change in the topology of the magnetic field can reduce the coupling of the disk with the surrounding media without changing of the magnetic flux through the disk.  Specifically, we consider the change from the configuration where magnetic field somewhat bends passing through the disk, i.e. a split monopole configuration to a dipole configuration as illustrated in Figure~\ref{fig:diag}. This change in the topology implies a decrease of the magnetic coupling between the disk and the surrounding medium, which in turn decreases the braking torque. The decrease is maximal when the magnetic field lines are closing outside the disk body as in Figure \ref{fig:diag}.  

\begin{figure}
\centering\includegraphics{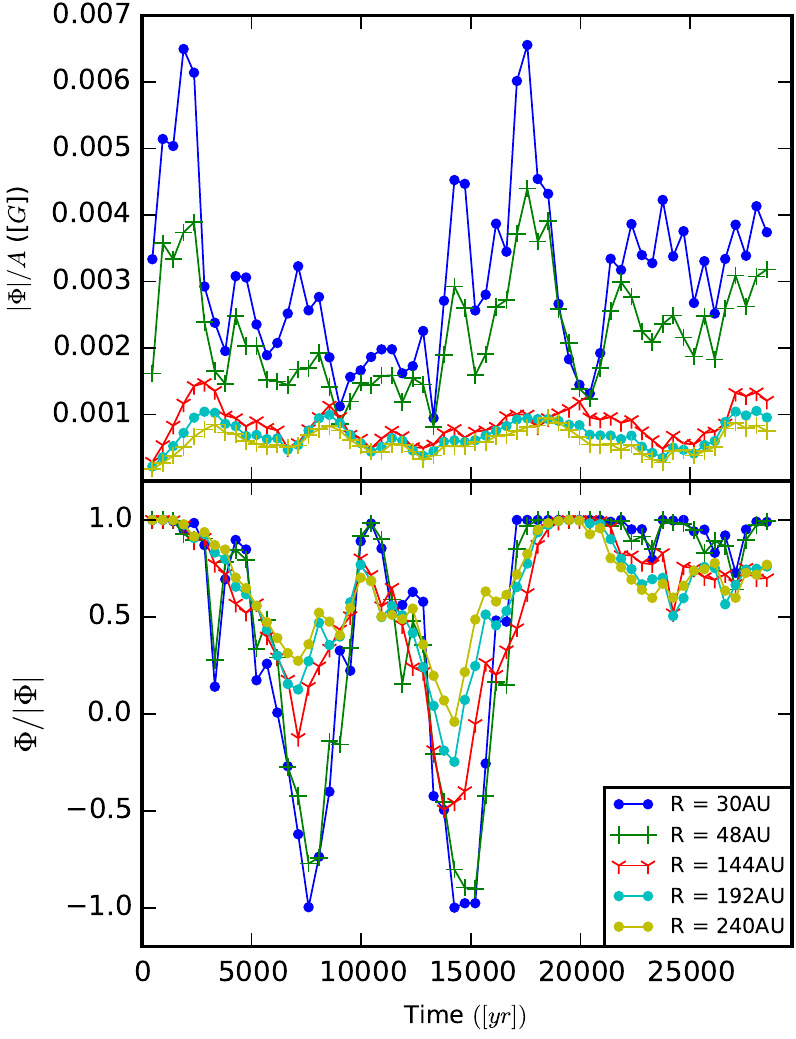}%[width=\linewidth,clip=true]
\caption{{\it Top panel:} absolute value of the magnetic flux, $|\Phi| = \int_s |B_z| dS$, normalized by the area of five 
concentric disk with different radii $R$ around the protostar in the plane 
of the disk. {\it Bottom panel:} ratio between the flux and the ``absolute flux'', inside concentric disks
with different radii $R$ in the plane of the disk.}
\label{fig:normflux}
\end{figure}

To measure the effects of the topology on the flux, we measure the magnetic flux, $\Phi = \int_S B_z dS$, and the absolute value of the magnetic flux, $|\Phi| = \int_s |B_z| dS$, crossing the area of concentric disks in the plane of the protostellar disk.  Magnetic field in a dipole configuration has field lines that cross twice the area of integration, one in the positive direction and one on the negative one (Figure \ref{fig:diag}). Hence the flux is going to be smaller than the absolute value of the flux. The ratio between the two fluxes reflects the presence of a dipole configuration (Figure~\ref{fig:normflux}). Furthermore, as the scale of the measurement increases, the effect of a magnetic flux reduction is enhanced due to the increase of the integration area, encompassing the flux with signal opposite from the flux crossing the interior of the disk.

There are three extreme events of reconnection where the formation of the dipole configuration is well observed during the evolution of the system. The times ($\sim$~7,500~yrs, $\sim$~15,000~yrs, and $\sim$~25,000~yrs) seem like consequence to the intensity peaks on the turbulence Figure~\ref{fig:sf}. Alternatively, there is another possibility that explains the reduction on the ratio of the fluxes in Figure~\ref{fig:normflux}, and this possibility is related to the transition to a superAlv\'{e}nic turbulence regime.  Since the simulation remains superAlv\'{e}nic for all times this is not possible.  More over, SuperAlv\'{e}nic turbulence produces a magnetic field wandering that can reduce the ratio of the fluxes. But for the latter possibility the decrease of the magnetic flux is not expected. We observe that the first two events corresponds to  transition to a dipole configuration while the last event can be attributed to the transition to superAlv\'{e}nic turbulence. Future work should study such events in more detail and for a wider set of scenarios.

In addition, the constant process of reconnection decreases the connection of the rotating disk with the envelope, as the magnetic field lines embedded in the disk get the ability of slippage in respect to the magnetic field lines embedded in the ambient interstellar gas. We plan to quantify these effects elsewhere. Here we just want to state that the ability of magnetic field to change its topology via reconnection opens another way of partial decoupling of magnetized disks from the conducting material outside the disks. 

\begin{figure}[]
\centering\includegraphics[width=\linewidth,clip=true]{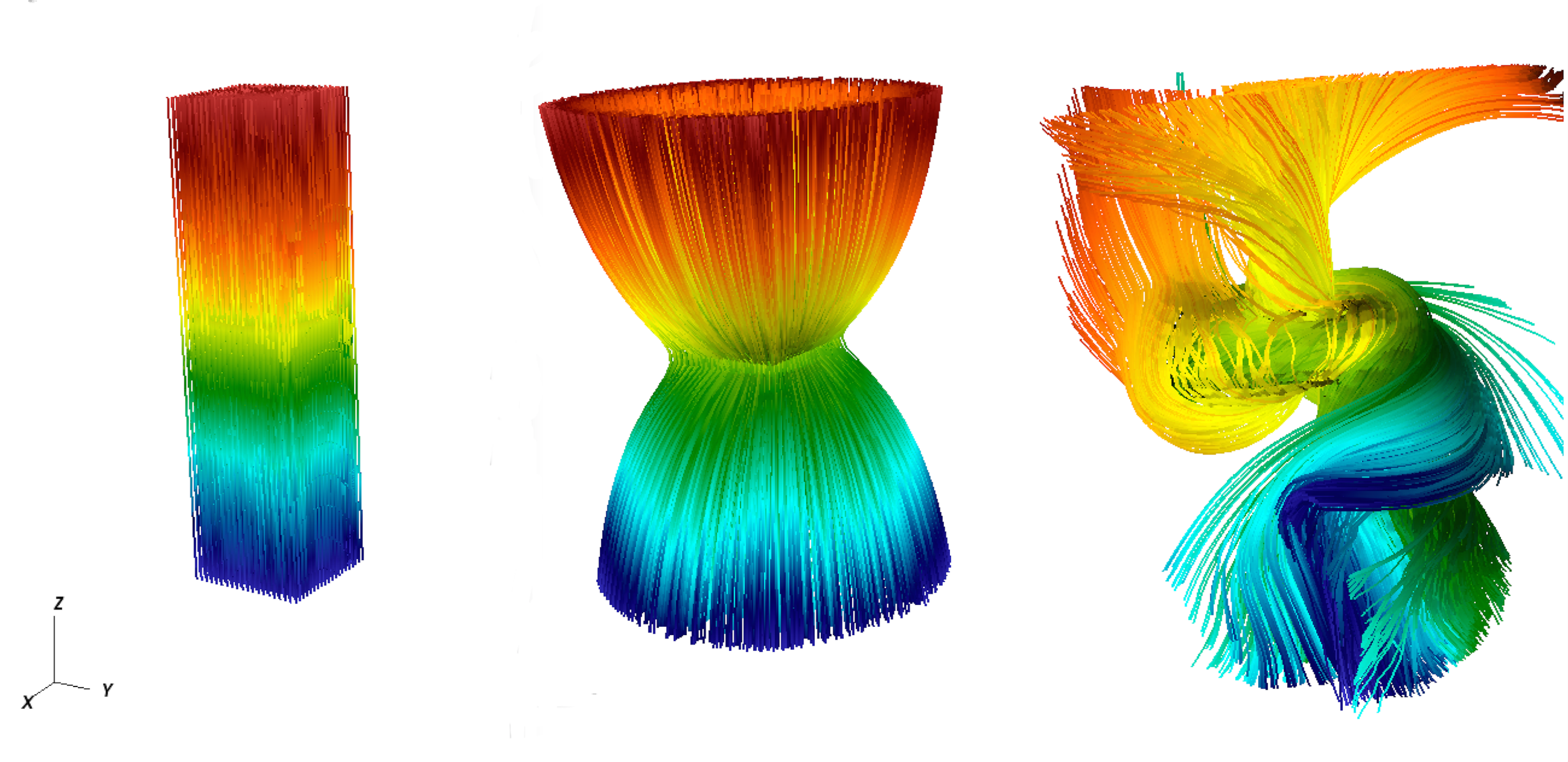}
\caption{Configuration of the magnetic field lines.  For all times, the seed particles used 
to visualise the magnetic lines are at the disk. {\it Right panel}: Magnetic field configuration 
at the initial time.  {\it Middle panel}: Magnetic field configuration at 2,000~yrs  {\it Left panel}: 
Magnetic field configuration at 14,000~yrs.  For all figures, the color represents the direction of 
the magnetic field lines.}
\label{fig:topo}
\end{figure}

As shown in Figure~\ref{fig:normflux}, there is a decrease in the absolute magnetic flux with time in the innermost circles. The gradual decrease in the flux --- even in the initial stages, where there is a build-up in the magnetic field intensity (Figure~\ref{fig:evol}) --- is interpreted as constant reconnection of the magnetic field to form a dipole configuration from the initial split monopole configuration. The change in the topology presented in Figure~\ref{fig:topo} shows the time-evolution of the magnetic field from a strict split monopole configuration to a complicated topology consisting of both split monopole and the dipole configurations.

%=======================================================
\section{Discussion} \label{sec:discussion}

We do not consider plasma effects as drivers for magnetic reconnection within this work. The irrelevance of small scale physics for turbulent reconnection was first shown in LV99. More detailed comparison of the effect of turbulent reconnection and Hall effect was given in \citet{eyink2011}. Finally, recent work by \citet{eyink2014} formulated a generalized Ohm's law which included contributions from standard Ohmic resistance, Hall resistance and other plasma effects to show the subdominance for reconnection of the plasma effects compared to the effects of turbulence. 

Other concepts, e.g. based on the so-called hyperresistivity \citet{bhattacharjee2003} were criticized as ill founded in our earlier papers \citep[see][]{lazarian2004, eyink2011, lazarian2015}. We refer our reader to the aforementioned publications.

Finally, we would like to stress that the process of turbulent reconnection is intrinsically different from the folklore concept of ``turbulent resistivity''. It is possible to show that ``turbulent resistivity'' description has fatal problems of inaccuracy and unreliability, due to its poor physical foundations for turbulent flow. It is true that coarse-graining the MHD equations by eliminating modes at scales smaller than some length $l$ will introduce a ``turbulent electric field'', i.e. an effective field acting on the large scales induced by motions of magnetized eddies at smaller scales. However, it is well-known in the fluid dynamics community that the resulting turbulent transport is not ``down-gradient'' and not well-represented by an enhanced diffusivity. The physical reason is that turbulence lacks the separation in scales to justify a simple ``eddy-resistivity'' description. As a consequence, energy is often not absorbed by the smaller eddies, but supplied by them, a phenomenon called ``backscatter''. In magnetic reconnection, the turbulent electric field often creates magnetic flux rather than destroys it.

We also point out that fast turbulent reconnection concept is definitely not equivalent to the dissipation of magnetic field by resistivity . While the parametrization of some particular effects of turbulent fluid may be achieved in models with different physics, e.g. of fluids with enormously enhanced resistivity, the difference in physics will inevitably result in other effects being wrongly represented by this effect. For instance, turbulence with fluid having resistivity corresponding to the value of ``turbulent resistivity'' must have magnetic field and fluid decoupled on most of its inertia range turbulent scale, i.e. the turbulence should not be affected by magnetic field in gross contradiction with theory, observations and numerical simulations. Magnetic helicity conservation which is essential for astrophysical dynamo should also be grossly violated. A more detailed discussion of the difference between the effects
of turbulent reconnection and the ill-founded idea of ``turbulent resistivity'' is given in
\citet{lazarian2015}.

Our study of does not include effects of ambipolar diffusion. We claim that for turbulent environments this is a subdominant effect. Our analysis of the reconnection diffusion in \citet{lazarian2015} shows that the process proceeds efficiently in the partially ionized gas and the difference between the reconnection diffusion within fully ionized gas and the partially ionized gas is insignificant on the scales larger than several Alfv\'{e}n turbulence damping scales. This is definitely true for the scales at which we consider reconnection diffusion in our paper.

%We also disregard effects of ``turbulent ambipolar diffusion'' \citep{fatuzzo2002, zweibel2002}. The latter effect was studied in 2D mixing with magnetic field perpendicular to the mixing plane \citep{heitsch2004}. We claim that that this sort of mixing is only possible in 2D, while in 3D any realistic mixing must include magnetic reconnection. If magnetic reconnection does not follow LV99 predictions, then the magnetic mixing is going to be constrained invalidating the predictions of the ``turbulent abmipolar diffusion'' theory. If, however, the reconnection proceeds as predicted by the turbulent reconnection theory, we deal with the reconnection diffusion process. 

The process that we invoked here is reconnection diffusion, which is different from "turbulent ambipolar diffusion" process discussed in a number of papers \citep{fatuzzo2002, zweibel2002}. It is discussed in detail in our earlier works  \citep[see][]{lazarian2014, lazarian2015d} why the processes required for the "turbulent ambipolar diffusion" are intrinsically dependent on fast turbulent reconnection and therefore can only take place in the presence of reconnection diffusion. As in turbulent fluids the ambipolar diffusion does not increase the rates of diffusion compared to reconnection diffusion we conclude \citep[see more in][]{lazarian2015d} that the concept of "turbulent ambipolar diffusion" is not useful and misleading.

%=========================================
\section{Conclusions} \label{sec:conclusion}

We measure the bulk magnetic field and angular momentum, as well as their evolution over time, for the process of disk formation.  From those measurements, it is easy to see that there are three different stages for the magnetic field and two for the angular momentum.  For the angular momentum, there is a build-up phase and then a steady-state one. The magnetic field has a build-up, decrease and steady-state faces. This decrease of the magnetic field implies that there should be a flux of the magnetic field out of the disk (associated with a topological reconfiguration) to reduce its field strength.  Because the bulk properties of the disk have to come from a measurable quantity from its borders, we measure the angular momentum flux through the three surfaces.  In all cases, it was found that the component driven by the magnetic field is much stronger than the one driven by the angular momentum one.  Its important to state that the dynamics are truly dominated by the magnetic field, and that this angular momentum flux modifies both the bulk magnetic field and angular momentum.

We found that the angular momentum flux due to the magnetic field is greater than the gas component, in correspondence with previous results by \citet{santos2010}.  Moreover, we found that the shearing due to the turbulent motions is not capable of compensating the angular momentum loss due to magnetic breaking and therefore such shearing cannot resolve the ``magnetic breaking catastrophe''. 

The angular momentum flux has three distinct regimens that matches with the ones of the bulk magnetic field, implying there is a magnetic diffusion.  Because we want to prove that reconnection diffusion is actually the mechanism responsible for the disk formation, we measure the properties of the turbulence using structure functions.  With that tool, we measure the power spectrum and the intensity of the turbulence.  We found that the intensity of the structure functions is related to the angular momentum flux, as it is expected from
the turbulent reconnection theory.  Hence, we conclude that reconnection diffusion is the mechanism that allows the dissipation of the magnetic field.

We also noticed that turbulent reconnection decreases the coupling of the rotating disk with the ambient interstellar matter. One of the processes is related  to the magnetic field changes from a split monopole to a dipole configuration. That change in itself implies that the connection of the disk with the ambient matter diminishes without the need of a decreasing magnetic field strength within the disk.  To test this idea, we measure the magnetic flux and its absolute value at the plane of the disk with different disk sizes. 

There are three instants where there is a drop on the ratio of fluxes.  This drops can correspond either to a change in the topology (for the first two events) or to a magnetic field wandering due to the super-Alfv\'{e}nic regimen (last event).  In each case, the drops corresponds to peaks in the turbulence intensity. These effects related to fast turbulent reconnection act to contribute to resolving the paradox of the ``magnetic breaking catastrophe'' but require further studies. 
%==========================================

\acknowledgments 
We thank Chris McKee, Dick Crutcher, Mordecai-Mark Mac Low, and Zach Pace for insightful discussions. AL and DFGC are supported by the NSF grant AST 1212096. Partial support for DFGC was provided by CONACyT (Mexico). AL acknowledges a distinguished visitor PVE/CAPES appointment at the Physics Graduate Program of the Federal University of Rio Grande do Norte and thanks the INCT INEspaço and Physics Graduate Program/UFRN, at Natal, for hospitality. RSL acknowledges support from a grant of the Brazilian Agency FAPESP (2013/15115-8). DFGC thanks the support of the Brazil-U.S. Physics Ph.D. Student and Post-doc Visitation Program.

\bibliographystyle{apj}
\bibliography{biblio}{}

\end{document}